# The reactivity of methanimine radical cation ($H_2CNH^{•+}$) and its isomer aminomethylene ($HCNH_2^{•+}$) with $C_2H_4$

D. Sundelin[a], D. Ascenzi[b], V. Richardson[b], C. Alcaraz[c,d], M. Polášek[e], C. Romanzin[c,d], R. Thissen[c,d], P. Tosi[b], J. Žabka[e] and W. Geppert[a] *

[a] Department of Physics, Stockholm University, Roslagstullsbacken 21, 10691 Stockholm, Sweden

[b] Department of Physics, University of Trento, Via Sommarive 14 – 38123 Italy

[c] Université Paris-Saclay, CNRS, Institut de Chimie Physique, UMR8000, 91405 Orsay, France

[d] Synchrotron SOLEIL, L'Orme des Merisiers, F-91192 Saint Aubin, Gif-sur-Yvette, France

[e] J. Heyrovsky Institute of Physical Chemistry of the Czech Academy of Sciences, Dolejškova 3, Prague 8, 18223 Czech Republic

* corresponding author: e-mail wgeppert@fysik.su.se phone: ++ 46 8 5537 8649

**Abstract**

Experimental and theoretical studies are presented on the reactivity of $H_2CNH^{•+}$ (methanimine) and $HCNH_2^{•+}$ (aminomethylene) isomers with ethylene. Selective isomer generation is performed via dissociative photoionization of apt neutral precursors and reactive cross sections as a function of photon and collision energies are measured. Main products for both isomers are H-elimination from covalently bound adduct (giving c-$CH_2CH_2CHNH^+$/$CH_2CHNHCH_2^+$) and H• atom transfer to yield $H_2CNH_2^+$. Differences in isomers' reactivity (cross sections and branching ratios) are discussed in light of calculations on reaction mechanisms. The astrochemical implications of the results are briefly addressed.

## 1. Introduction

Ion-induced processes have been regarded for a long time to play a pivotal role in the synthesis of complex molecular species in the interstellar medium and atmospheres of planets and their satellites[1]. More and more complex ions have been detected using state-of-the-art terrestrial single-dish telescopes and interferometers. With the increasing complexity of species, the role of isomers becomes more crucial. The existence of isomeric ions in the interstellar medium has long been known and, in the case of $HCO^+$ and HOC[+] their relative abundance ratio has been assessed[2], which can allow conclusions about the chemistry of their environments. Isomers have different spectroscopic and chemical processes and, since isomerisation barriers for many isomeric species are too high to be overcome thermally under interstellar conditions, they have to be treated as different species with distinctive chemical reactivity, which is important for model calculations. Whereas isomers of ions often possess different dipole moments and other spectroscopic properties and thus can be distinguished in radioastronomic observations [2], assessment of ion abundances in planetary and satellite atmospheres often relies on mass spectrometers which cannot distinguish between isomers. Thus their relative abundance often remains elusive in those environments. However, the novel Atacama Large Interferometer Array (ALMA) allowed to obtain column density information as a function of altitude for HCN and HNC[3]. The unprecedented resolution of ALMA will also enable to retrieve data about the abundance and distribution of isomeric ions in different astronomic environments. In order to perform model calculations rationalising these abundances, it is important to assess the reactivity of ionic isomers with common interstellar and atmospheric molecules. Several efforts have been made during the last years to accomplish this task using apt precursor molecules to produce one of the isomers purely or preferentially[4]. Especially VUV photoionisation has proved a versatile tool for selective production of isomers[5]. Noble gas tagging has been successfully employed to characterise isomeric ions and to determine their relative abundance after electron ionisation using different neutral precursors[6].

Titan is one of the most interesting objects in the solar system, since, as our own planet, it possesses a dense nitrogen-dominated atmosphere[7]. The ion and neutral mass-spectrometer (INMS) onboard the Cassini spacecraft dicovered that Titan's atmosphere is one of the most complex in the solar system, containing large hydrocarbons and nitrogen-bearing compounds[8, 9]. This very versatile chemistry probably mainly starts through ionization and/or dissociation of its main components $N_2$ and $CH_4$ by extreme ultraviolet (EUV) radiation[10] or magnetospheric electrons. Primarily formed ions and radicals can subsequently undergo ion-neutral and radical-neutral reactions leading to more complex compounds[11, 12]. Especially ion-neutral pathways involving unsaturated hydrocarbons (*e.g.* acetylene and ethylene) as building blocks are frequently held responsible for the production of complex ions detected in Titan's atmosphere[12, 13]. Positive ions with masses up to 99 amu were detected by INMS at altitudes of 950 km above the surface and the presence of cations with higher masses (up to 350 amu) were observed with the Cassini plasma spectrometer ion beam sensor (CAPS-IBS)[14], Thus, it is important to elucidate the production routes of these species. Reactions of ionic nitrogen-containing compounds with hydrocarbons could play a decisive role in that. However, many of these species are protonated nitriles which are comparatively unreactive and mostly destroyed by dissociative recombination in Titan's atmosphere[15]. But this does not hold for ($H_2CNH^{\bullet+}$) and its isomer aminomethylene ($HCNH_2^{\bullet+}$) which are reactive radical cations and could act as a template for forming larger neutral molecules through chain elongation reactions with unsaturated and saturated hydrocarbons and subsequent dissociative recombination of the resulting enlarged cation. Although the mass signal at *m/z* 29 recorded by INMS can be regarded to be mainly due to a mixture of $C_2H_5^+$ and $H^{13}CNH^+$ ions, model calculations predict a density of $HCNH_2^{•+}$ and its isomers amounting to $1.1 \times 10^{-2}$ cm$^{-3}$ in Titan's ionosphere[15]. Thus, such species may play a certain role in building up larger nitrogen-containing chemicals which can then further react to form aerosols associated to Titan's orange-coloured haze. This paper presents a reactivity study of $H_2CNH^{•+}$ and its isomer $HCNH_2^{•+}$ with ethylene using synchrotron radiation and apt precursors to generate the charged species selectively.

## 2. Methodology

The experimental and theoretical methodologies used have already been decribed in details in an accompanying paper to this Special Issue devoted to reactivity of the title ions with $C_2H_4$ [16] and in Ref. [17], so only the briefest summaries is given here.

### *2.1 Experimental Set-Up*

Experiments were performed using the CERISES apparatus[17,18], at the DESIRS beamline[19] of the SOLEIL synchrotron radiation facility. CERISES is a guided ion beam tandem mass spectrometer composed of two octopoles located between two quadrupole mass filters. $[H_3CN]^{•+}$ isomers were produced by dissociative photoionization (in the photon energy range 9.5÷13.5 eV) of gaseous precursors (see **Section 2.3**) introduced at roughly $10^{-6}$ mbar in the ion source. Ethylene was introduced in the reaction cell (surrounding the first octopole) at a dynamic pressure of 2.6 x$10^{-7}$ bar, which guarantees operation in single collision mode, reduces secondary reactions and limits the parent ion attenuation to less than 10%. Collision energies were set by the difference between the ion source and the reaction cell potentials and the retarding potential method[20] was used to measure a parent beam FWHM of 400 mV (0.196 eV in the centre-of-mass frame). By changing the potentials of the reaction cell and all subsequent elements up to 20V, the collision energy can be increased up to 9.82 eV in the centre-of-mass frame.

### *2.2 Theoretical Methodology*

The mechanisms for the reactions of both isomers were studied using GAUSSIAN[21]. Intermediate structures were calculated at the MP2/6-31G*, MP2/6-311++G** and MP2/cc-pVTZ level and the identities of transition states and minima were checked by frequency calculations. Zero-point energy corrections were employed to the obtained energies. IRC calculations were performed at the MP2/6-31G* level to ensure that the transition states connect the correct minima. The dissociation pathways were probed through scan computations along the dissociation coordinates of the corresponding minima. Energy calculations were carried out for all stationary points at the CCSD(T)/6-311++G**and CCSD(T)/cc-pVTZ level with zero-point energy corrections taken from the MP2/6-311++G** and MP2/cc-pVTZ levels, respectively. Full results are given in the **Supplementary Information.**

## 2.3 Ions generation and checks on impurities

The generation of the two ions through dissociative photoionization of different neutral precursors is described elsewhere[16, 22]. Due to lack of space, the reader is referred to such references and to the **Supplementary Information** of this work for a detailed discussion of the potential impact of isobaric impurities ($C_2H_5^+$, $H^{13}CNH^+$ and $^{13}CCH_4^{\bullet+}$) on the reaction with $C_2H_4$. Only the main conclusions are reported here.

The $HCNH_2^{\bullet+}$ or $H_2CNH^{\bullet+}$ isomers are generated from dissociative photoionization of cyclopropylamine ($c$-$C_3H_5NH_2$) or azetidine (c-$CH_2CH_2CH_2NH$)[23, 24, 25], respectively. In both cases the experimental appearance energies of fragments at *m/z* 29 are 10.2±0.1 eV[16]. Theoretical calculations, control experiments performed by mass selecting parent ions at *m/z* 28 from both azetidine and cyclopropylamine, as well as literature data on the reactivity of $C_2H_5^+$, $C_2H_4^+$ and $HCNH^+$ with $C_2H_4$ [26] allowed us to conclude that:

a) dissociative photoionization of cyclopropylamine at photon energies above ~12.4 eV leads to contamination from $C_2H_5^+$ which can react with $C_2H_4$ to give $C_3H_5^+$ (*m/z* 41). This contamination channel seems less relevant in the azetidine case.
b) although the production of $HCNH^+$ from dissociative photoionization of both precursors has similar appearance energies as $HCNH_2^{\bullet+}/H_2CNH^{\bullet+}$ [16], the reaction of $HCNH^+$ with $C_2H_4$ being endothermic and leading exclusively to the proton transfer product $C_2H_5^+$, we expect any contamination from $H^{13}CNH^+$ in the reagent beam not to interfere with the results discussed in the following.
c) dissociative photoionization of azetidine at photon energies above ~11.3 eV leads to a small contamination from $C_2H_4^{\bullet+}$ which can react with $C_2H_4$ to give $C_3H_5^+$ (*m/z* 41) plus a minor amount of $C_4H_7^+$ (*m/z* 55). However, since the ratio of the photodissociation yields for *m/z* 28 and 29 at the explored photon energies is in the range 0.3÷1.0, we expect a small contamination of $^{13}CCH_4^{\bullet+}$ in the reagent beam, which may be responsible of the small signal observed at *m/z* 41 but does not interfere with the results presented here.

## 3. Results and discussion: reactivity experiment

The reaction of both isomers yielded products at *m/z* 30, 42 and 56 with the reaction of $H_2CNH^{\bullet+}$ also showing a product at *m/z* 28 significant enough for it to be separated from the parent signal. The reaction pathways for these processes (Reactions **1-4c**) with the relevant reaction enthalpies estimated from literature values are summarized in **Table 1**.

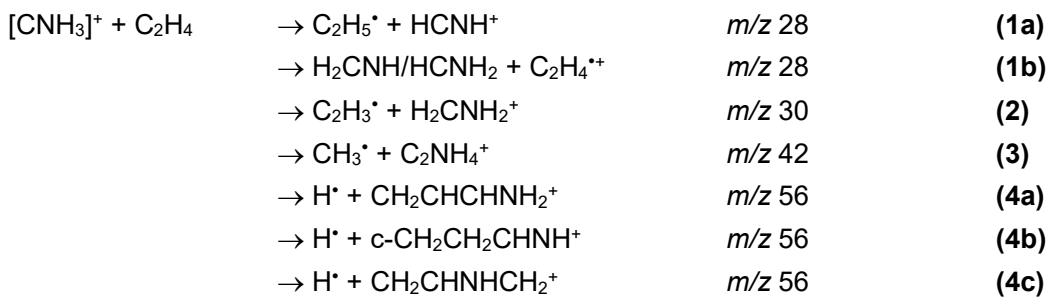

| [CNH$_3$]$^+$ + C$_2$H$_4$ | → C$_2$H$_5^{\bullet}$ + HCNH$^+$ | *m/z* 28 | **(1a)** |
| | → H$_2$CNH/HCNH$_2$ + C$_2$H$_4^{\bullet+}$ | *m/z* 28 | **(1b)** |
| | → C$_2$H$_3^{\bullet}$ + H$_2$CNH$_2^+$ | *m/z* 30 | **(2)** |
| | → CH$_3^{\bullet}$ + C$_2$NH$_4^+$ | *m/z* 42 | **(3)** |
| | → H$^{\bullet}$ + CH$_2$CHCHNH$_2^+$ | *m/z* 56 | **(4a)** |
| | → H$^{\bullet}$ + c-CH$_2$CH$_2$CHNH$^+$ | *m/z* 56 | **(4b)** |
| | → H$^{\bullet}$ + CH$_2$CHNHCH$_2^+$ | *m/z* 56 | **(4c)** |

**Table 1.** Reaction enthalpies for the reaction: $HCNH_2^{•+}/H_2CNH^{•+} + C_2H_4 \rightarrow$ products

| Products | Reaction | $\Delta H°$ with $HCNH_2^{•+}$ (eV) [a] | $\Delta H°$ with $H_2CNH^{•+}$ (eV) [a] |
|---|---|---|---|
| $C_2H_5^• + HCNH^{•+}$ | 1a | -0.11[b] | -0.27[b] |
| $H_2CNH/ HCNH_2 + C_2H_4^{•+}$ | 1b | +2.31[c] | +0.58[c] |
| $C_2H_3^• + H_2CNH_2^+$ | 2 | -0.40[d] | -0.56[d] |
| $CH_3^• + C_2NH_4^+$ | 3 | -1.16[e] / +0.14[f] | -1.32[e] /-0.03[f] |
| $H^• + CH_2CHCHNH_2^+$ | 4a | -1.21[g] | -1.37[g] |
| $H^• + c\text{-}CH_2CH_2CHNH^+$ | 4b | -0.61[g] | -0.77[g] |
| $H^• + CH_2CHNHCH_2^+$ | 4c | -0.65[g] | -0.81[g] |

[a] All values in the Table have been evaluated using $\Delta_fH°(HCNH_2^{•+})$=10.67 eV [27] and $\Delta_fH°(H_2CNH^{•+})$=10.84 eV [27]

[b] Assuming the formation of $HCNH^+$ ($\Delta_fH°$ = 9.87 eV[28]) plus the ethyl radical $C_2H_5$ with $\Delta_fH°$= 1.23 eV[28]

[c] Assuming the formation of neutral methanimine ($\Delta_fH°$ = 0.91 eV[29]) or aminomethylene ($\Delta_fH°$ = 2.47 eV from calculations[31]

[d] Assuming the formation of the vinyl radical ($C_2H_3$) with $\Delta_fH°$ = 3.1 eV[30] and aminomethylium ion $H_2CNH_2^+$ with $\Delta_fH°$ = 7.72 eV[30]. This pathway is also predicted by our calculations to lead to the product ion with $m/z$ 30

[e] Assuming the formation of the most stable of the $C_2NH_4^+$ isomers, *i.e.* protonated acetonitrile $\Delta_fH°$= 8.55 eV[28].

[f] Assuming the formation of protonated ketenimine ($CH_2CNH_2^+$ $\Delta_fH°$ = 9.85 eV [28]), as predicted by present calculations (see **Sec. 4**).

[g] Assuming estimates from Ref. [28] for $\Delta_fH°$ of the various $C_3H_6N^+$ isomers, namely $\Delta_fH°$ = 7.75 eV for the $CH_2CHCHNH_2^+$ isomer, $\Delta_fH°$ = 8.31 eV for the $CH_2CHNHCH_2^+$ isomer and $\Delta_fH°$ = 8.35 eV for the c-$CH_2CH_2CHNH^+$ isomer.

The reaction with $C_2H_4$ has only previously been studied with $HCNH_2^{•+}$ by FT-ICR, which lead to adduct formation at $m/z$ 57 plus various loss channels: of $C_2H_5^•$ (channel **1a**), of $NH_2^•$ (to yield $C_3H_5^+$, $m/z$ 41), and of $H^•$ (channels **4a-b-c**)[25], with the latter being about three times more abundant than the others. This same work also examined the reaction with $C_2D_4$ wich showed a 27:73 ratio for the loss of H:D from the adduct equivalent to reactions **4a-c**. In our study the production of an ion at $m/z$ 41 ($C_3H_5^+$) from $HCNH_2^+$ generated from cyclopropylamine is observed only at photon energies above ~12.4 eV and it is attributed to reactions of $C_2H_5^+$ contaminants (see **Supplementary Information**). It is likely that this is also the case when $HCNH_2^+$ is formed starting from cyclopropylamine in an electron ionization source[25].

Cross sections (CSs) for the reaction of both ions as a function of the photon energy ($E_{phot}$) at fixed collision energy ($E_{CM}$) are given in **Fig. 1**, while in **Fig. 2**. CSs as a function of the collision energy are shown. Branching ratios (b.r.) are given in **Table 2**.

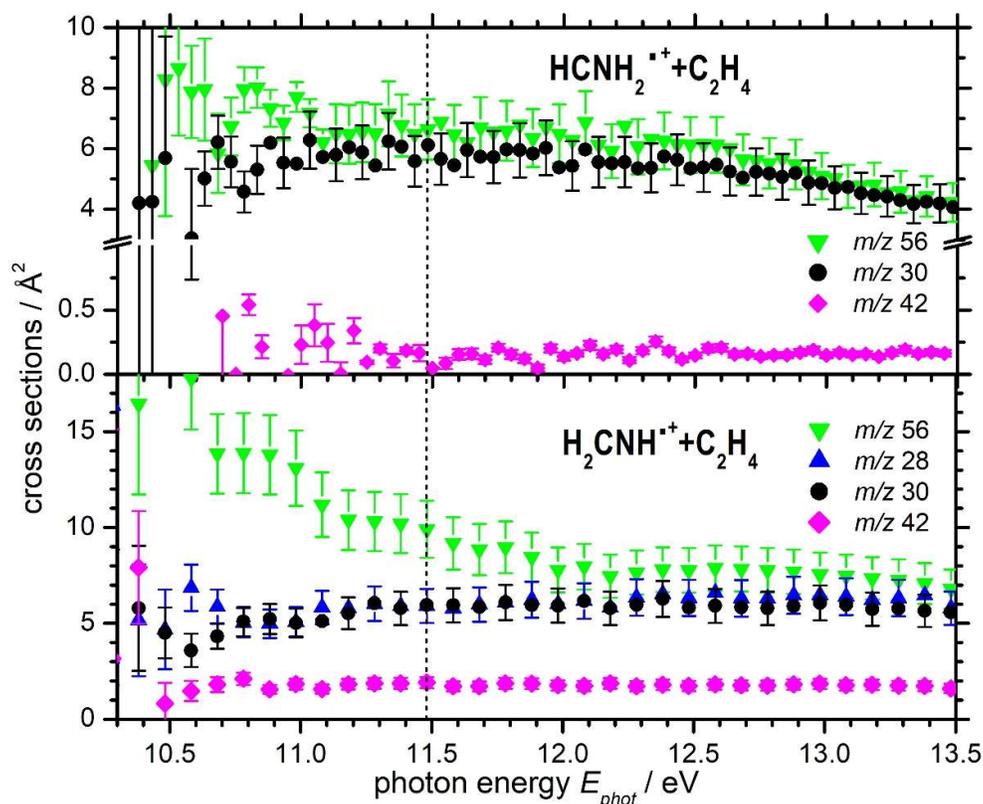

**Fig. 1**: Reactive cross sections as a function of $E_{phot}$ for the reaction of $HCNH_2^{\cdot+}$ (top) and $H_2CNH^{\cdot+}$ (bottom) with $C_2H_4$. Collision energies are $E_{CM}=0.15$ eV (top) and $E_{CM}=0.13$ eV (bottom). The vertical dashed lines indicate the photon energy at which data have been collected as a function of $E_{CM}$ (see **Fig. 2**)

**Table 2.** Branching ratios (b.r.) for the reaction of $HCNH_2^{\cdot+}$ and $H_2CNH^{\cdot+}$ with $C_2H_4$

| product (m/z) | b.r. $HCNH_2^{\cdot+}$ [a] | b.r. $H_2CNH^{\cdot+}$ [a] |
|---|---|---|
| 28 | 0.00 ± 0.01 | 0.24 ± 0.02 |
| 30 | 0.45 ± 0.04 | 0.23 ± 0.02 |
| 42 | 0.010 ± 0.005 | 0.08 ± 0.02 |
| 56 | 0.54 ± 0.04 | 0.44 ± 0.02 |

[a] Extracted from data at $E_{phot}$ = 11.48 eV and $E_{CM}$ = 0.13 eV

The *m/z* 56 product is the dominant one at all photon energies for both isomers, indicative of a high level of adduct formation. For the $HCNH_2^{\cdot+}$ isomer this channel shows no significant trends as a function of the photon energy indicating that it proceeds independently of the internal energy of the reactant ion. The decrease in CS at $E_{phot}$ > 12.4 eV, also observed for the other major product at *m/z* 30, is an artefact due to the increasing contamination of $C_2H_5^+$ in the parent beam (see discussion in the **Supplementary Information**), which reduces the relative amount of $HCNH_2^{\cdot+}$ ion available for reaction.

The next most significant product for the HCNH$_2^{•+}$ isomer is at *m/z* 30, with similar trend with $E_{phot}$ as *m/z* 56. A minor channel at *m/z* 42 is detected and approximately constant (within uncertainty) as a function of the photon energy.

For the H$_2$CNH$^{•+}$ isomer, after the *m/z* 56 channel, the *m/z* 28 and 30 are the next most significant pathways, with equivalent CSs over the entire photon energy range. It is worth noting that at low photon energies the *m/z* 56 channel shows initial decrease before levelling off above 12 eV, whereas the *m/z* 30 channel undergo a slight increase over this same photon energy range. This is discussed further in **Section 5**. The minor channel at *m/z* 42 is also observed, being largely constant but notably more intense than for the other isomer.

In terms of trends as a function of the collision energy (**Fig. 2**), they are very similar for the *m/z* 30 from both isomers, with a gradual decrease with increasing $E_{CM}$ indicative of a barrierless exothermic process. Products at *m/z* 56 from both isomers show a slightly sharper decrease with increasing collision energy, indicative of a mechanism proceeding via adduct formation. This decrease is notably sharper for the H$_2$CNH$^{•+}$ isomer (see **Discussion** for a possible explanation). The *m/z* 42 channel also shows a similar trend for both isomers, with a predominant low collision energy peak typical of a barrierless mechanism with a secondary slight rise at higher collision energies, indicating a second process having an energy barrier. The *m/z* 28 product, only observed with H$_2$CNH$^{•+}$, shows an initial decrease with increasing collision energy indicative of a barrierless process, but then displays a significant rise to a plateau. This is a strong indication of the combination of a barrierless channel with either an endothermic pathway or one having an energy barrier. A detailed discussion of the trends of all products in light of computational results is presented in **Section 5**.

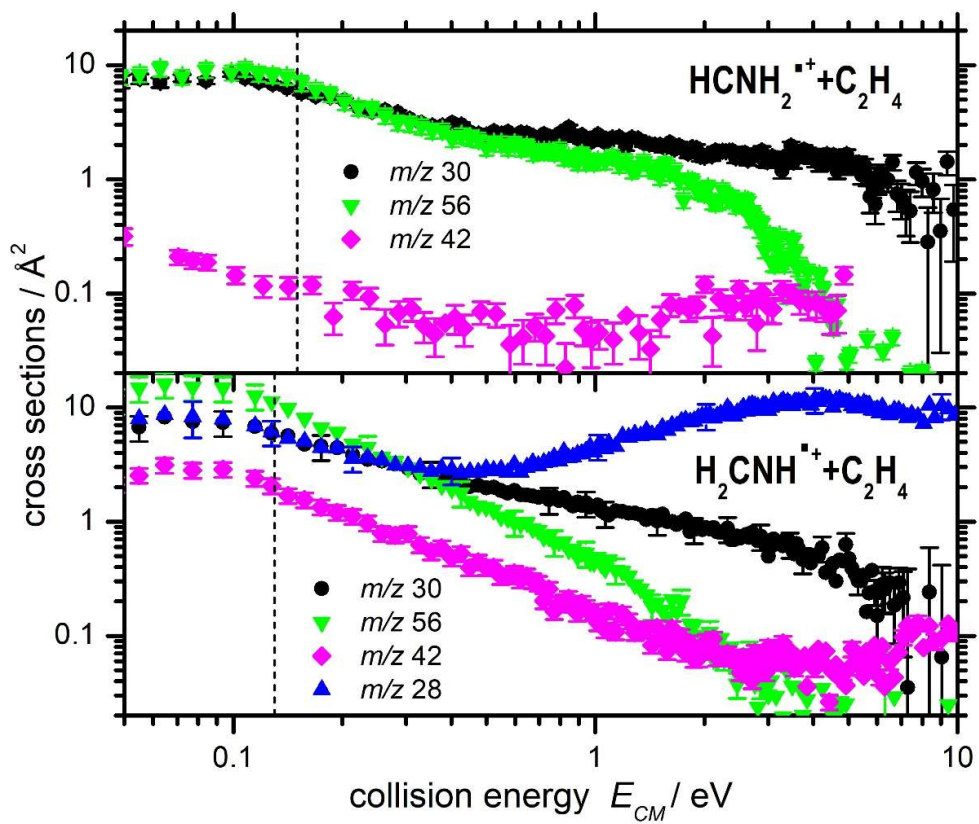

**Fig. 2**: Reactive cross sections as a function of the collision energy $E_{cm}$ for the reaction of HCNH$_2^{•+}$ (top) and H$_2$CNH$^{•+}$ (bottom) with C$_2$H$_4$. The photon energy is fixed at $E_{phot}$=11.48 eV, with the exception of m/z 42 from HCNH$_2^{•+}$ for which is equal to 12.48 eV. The vertical dashed lines indicate the collision energy at which data have been collected as a function of $E_{phot}$ (see **Fig. 1**)

### 4. Computational Results

In agreement with previous theoretical studies,[27,32] our calculations show that the aminocarbene (HCNH$_2^{•+}$) is more stable than the methanimine radical cation (H$_2$CNH$^{•+}$) by about 17.9 kJ·mol$^{-1}$ and the two isomers are separated by an isomerisation barrier for 1,2-H shift of ~280 kJ·mol$^{-1}$ (with respect to H$_2$CNH$^{•+}$ energy). This value is substantially higher than the barrier associated to H loss from either isomers, thus rendering dissociation favoured over isomerization.

Both isomers are able to form either van der Waals complexes or covalently-bound adducts with ethylene. In each case, the van der Waals complex results from the approach of the ion to the neutral by the non-radical terminus, whereas the covalent adducts are formed by approach to the neutral by the radical end. Note that the relative enthalpies reported in the following are always given (in kJ·mol$^{-1}$) with respect to the sum of the energies of the separated reactants H$_2$CNH$^{•+}$ and C$_2$H$_4$.

**HCNH$_2^{•+}$ isomer** (pathways are described graphically in **Fig. 3** and **4**): formation of the covalently-bound adduct **A12** (-245.6) proceeds via the van der Waals complex **A11** (- 65.0) and the transition state **TS12** (-40.7 kJ/mol). **A12** can readily eject a H via **TS13** (-98.9) to give the *m/z* 56 product CH$_2$CHCHNH$_2$$^+$ (-130.8). The other major product channel is the *m/z* 30 one, which can be formed in two ways: from **A12** which undergoes an isomerization to **A18** (-172.9) via **TS17** (-139.2) prior to a barrierless fragmentation into H$_2$CNH$_2$$^+$ and C$_2$H$_3$$^•$ (-45.9), or via a van der Waals cluster of the reactants **A21** (-63.8), which can transfer an H atom to give the cluster **A10** (-88.4) via **TS21** (-18.3). This species can then barrierlessly fragment to the same products.

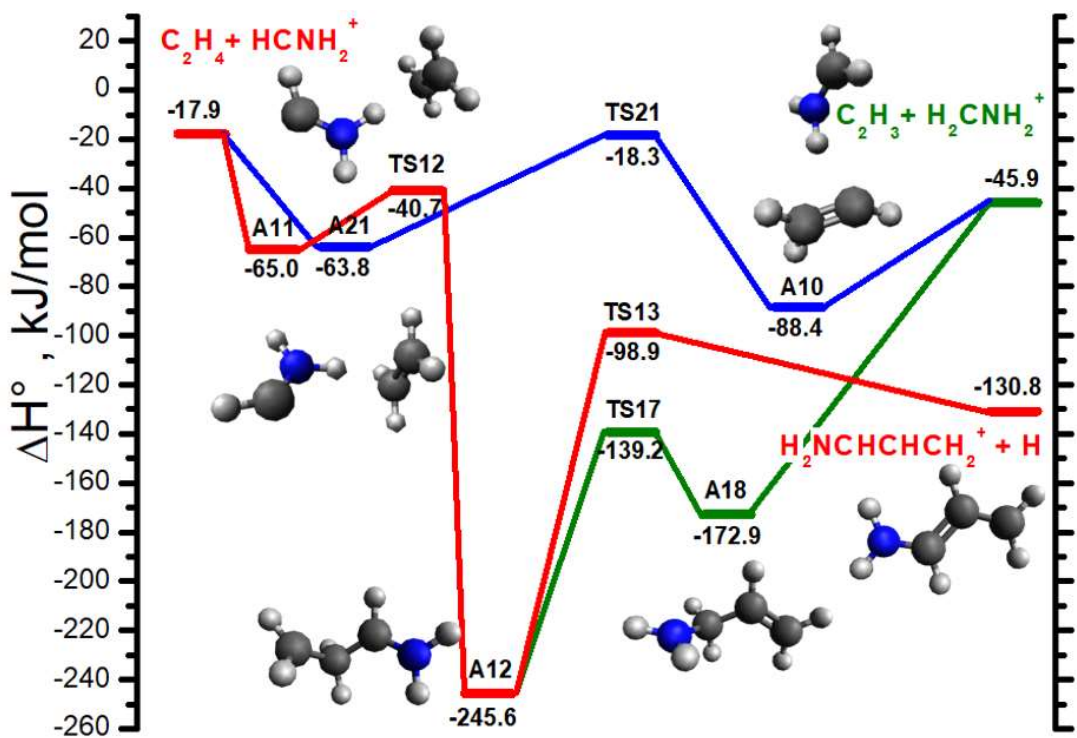

**Fig. 3**. *HCNH$_2$$^{•+}$ plus C$_2$H$_4$*: relative enthalpies and reaction pathways leading to products H$_2$CNH$_2$$^+$ plus C$_2$H$_3$$^•$ (in blue and green) and NH$_2$CHCHCH$_2$$^+$ plus H$^•$ (in red). The zero value for enthalpies (in kJ·mol$^{-1}$) corresponds to the separated H$_2$CNH$^{•+}$ plus C$_2$H$_4$ reactants. Calculations are at the UCCSD(T)/cc-pVTZ level of theory.

The formation of the *m/z* 42 product from HCNH$_2$$^{•+}$ can proceed from **A12** via **TS18** (-82.7) to give **A15** (-201.6). This can rearrange to **A16** (-206.5) via **TS19** (-199.4). **A16** can then transform, via **TS20** (-51.4), to **A17** (-81.1) which subsequently fragments barrierlessly to give CH$_2$CNH$_2$$^+$ plus CH$_3$$^•$ (-69.1). Although we have not measured the CSs for the *m/z* 28 product from this isomer as the signal was too minor to make it separatable from the parent beam, a submerged pathway for its formation is here presented. **A12** can convert into its rotamer **A8** (-251.3) via **TS6** (-234.4). **A8** can undergo a [1,4] H shift to give **A13** (-166.4) via **TS14** (-126.4). **A13** can then convert into the rotamer **A14** (-166.5) via **TS15** (-164.7). In turn, **A14** can cleave the central C-C bond to form **A5** (-66.5), a complex of HCNH⁺ and C₂H₅•, via **TS16** (-67.2). This complex can barrierlessly fragment to give products (-39.6).

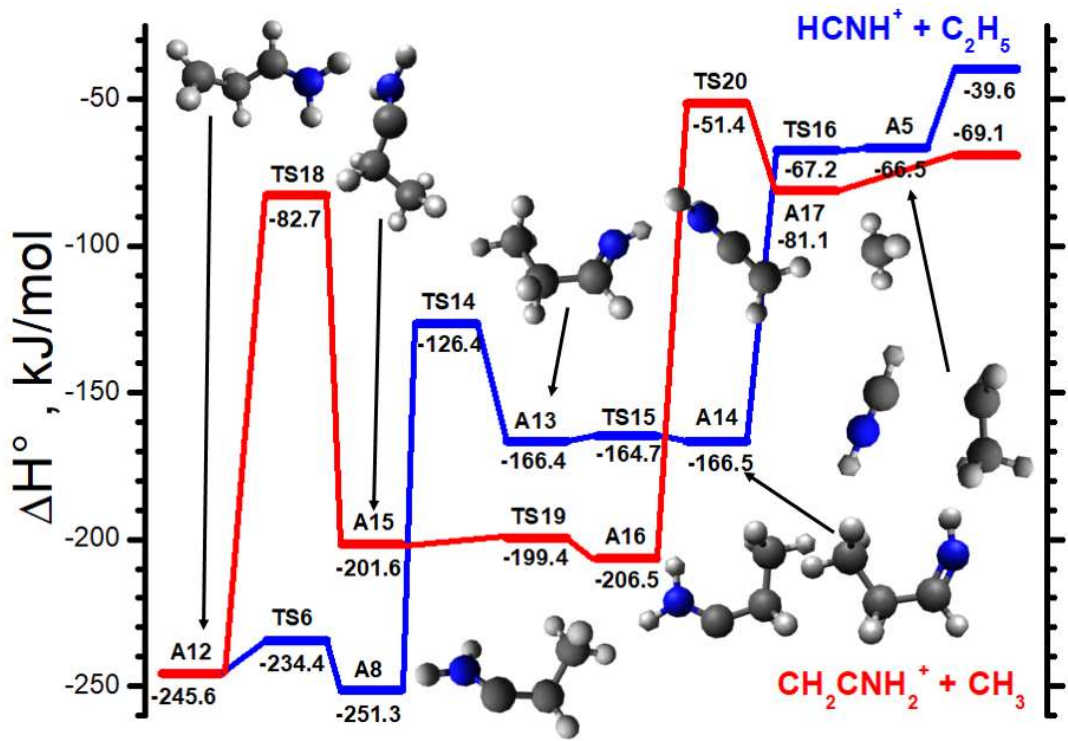

**Fig. 4**. *HCNH₂•⁺ plus C₂H₄*: relative enthalpies and reaction pathways leading to products HCNH⁺ plus C₂H₅• (in blue) and CH₂CNH₂⁺ plus CH₃• (in red). The zero value for enthalpies (in kJ·mol⁻¹) corresponds to the separated H₂CNH•⁺ plus C₂H₄ reactants. Calculations are at the UCCSD(T)/cc-pVTZ level of theory.

**H₂CNH•⁺ isomer** (pathways are described graphically in **Fig. 5** and **6**): the covalently bound adduct **A1** (-210.7) is formed barrierlessly. From **A1** two pathways are possible to form different isomeric products at *m/z* 56. In one case **A1** can rearrange into its rotamer **A2** (-210.5) via **TS1** (-202.5). This can then eject an H to form CH₂CHNHCH₂⁺ (-76.3) via **TS2** (-31.4). An alternative pathway proceeds through **A1** cyclisation to give **A3** (-189.6) via **TS3** (-98.3). This can then eject a H to give c-CH₂CH₂CHNH⁺ (-69.2) via **TS4** (-48.3).

This isomer also has two possible mechanisms for formation of products at *m/z* 30 amu, though only one of which has fully submerged energy barriers. This pathway proceeds via the van der Waals complex **A9** (-20.1) and involves an H transfer from C₂H₄ to the ion to give **A10** (-88.4) via **TS11** (-34.0). **A10** can barrierlessly separate to form H₂CNH₂⁺ plus C₂H₃• (-45.9). The other mechanism requires the rearrangement of **A2** into **A22** (-164.0), an isomerisation that is hampered by a large barrier going via **TS10** (+28.0). Subsequently, the barrierless rupture of the C-N bond in **A8** leads to products.

The formation of the other major product at *m/z* 28 first involves a [1,4] H shift from **A1** to give **A4** (-147.7) via **TS5** (-96.8). **A4** can then rearrange to form its rotamer **A19** (-148.0) via **TS22** (-116.5). This can then cleave the central C-N bond to give **A20** (-99.0), a van der Waal complex of HCNH$^+$ and C$_2$H$_5$· via **TS23** (-35.3). **A20** can then separate barrierlessly to give products HCNH$^+$ plus C$_2$H$_5$· at -39.6 kJ·mol$^{-1}$.

The mechanism for the formation of *m/z* 42 product proceeds from **A2** via **TS7** (-9.7) to give **A6** (-169.6). This can then transform to its rotamer **A7** (-172.0) via **TS8** (-169.0) which in turn can fragment via **TS9** (-65.8) to give CH$_2$CNH$_2$$^+$ and CH$_3$· (-67.1).

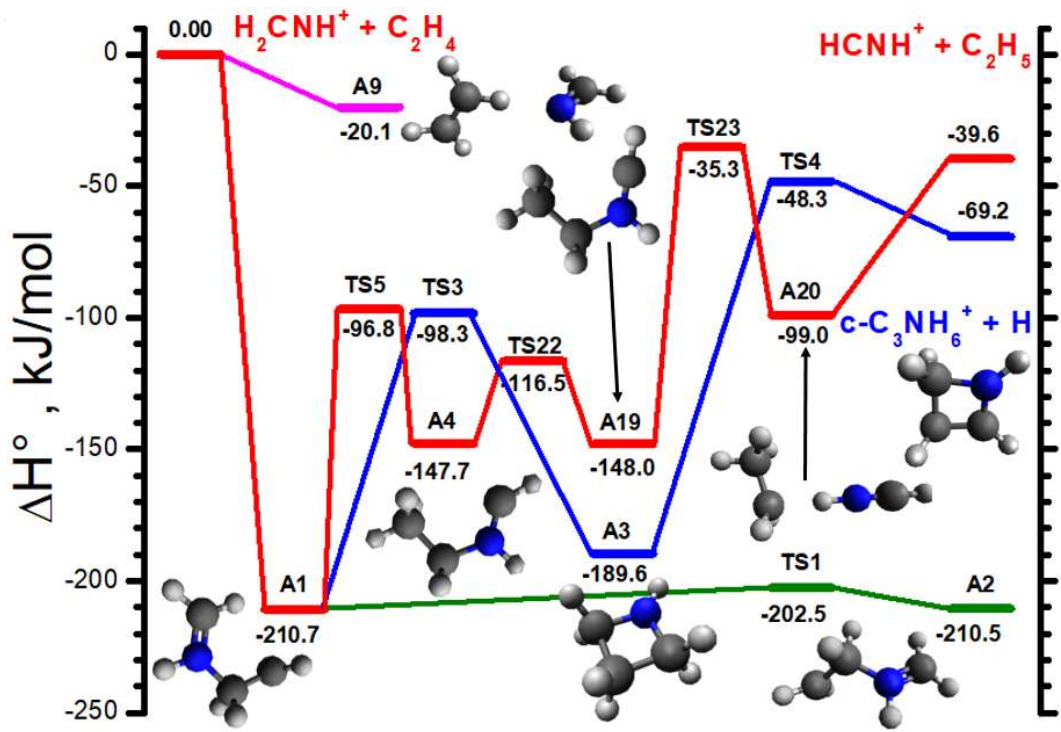

**Fig. 5**. *H$_2$CNH$^{·+}$ plus C$_2$H$_4$*: relative enthalpies and reaction pathways leading to products HCNH$^+$ plus C$_2$H$_5$· (in red) and c-CH$_2$CH$_2$CHNH$^+$ plus H· (in blue). The zero value for enthalpies (in kJ·mol$^{-1}$) corresponds to the separated H$_2$CNH$^{·+}$ plus C$_2$H$_4$ reactants. Calculations are at the UCCSD(T)/cc-pVTZ level of theory.

## 5. Discussion

The chemistry of HCNH$_2$$^{·+}$ is dominated by processes involving the formation of intermediate adducts, with all three main channels proceeding at least partly in this way. The dominance of the *m/z* 56 channel at low collision energies can be explained by the fact that this is the most exothermic channel and only involves a single rearrangement (from the initial adduct **A11** to **A12**) for formation. Hence, it is the most favoured channel both kinetically and thermodynamically.

For H$_2$CNH$^{·+}$ the decrease of the *m/z* 56 product with increasing collision energy is more pronounced. This difference can be attributed to the fact that the lowest energy pathway proceeds via a cyclisation, which is likely to be inhibited by increasing the collision energies; whereas this is not the case for HCNH$_2$$^{·+}$ where a direct ejection pathway is available. A similar explanation holds for the slight decrease in CS as the photon energy (hence internal energy of the H$_2$CNH$^{·+}$ reactant) is increased, since internal energy inhibits the cyclisation, and while there is a direct ejection pathway with submerged barriers (see **Sec. 4**), this involves higher barriers than the pathway via the cyclised adduct.

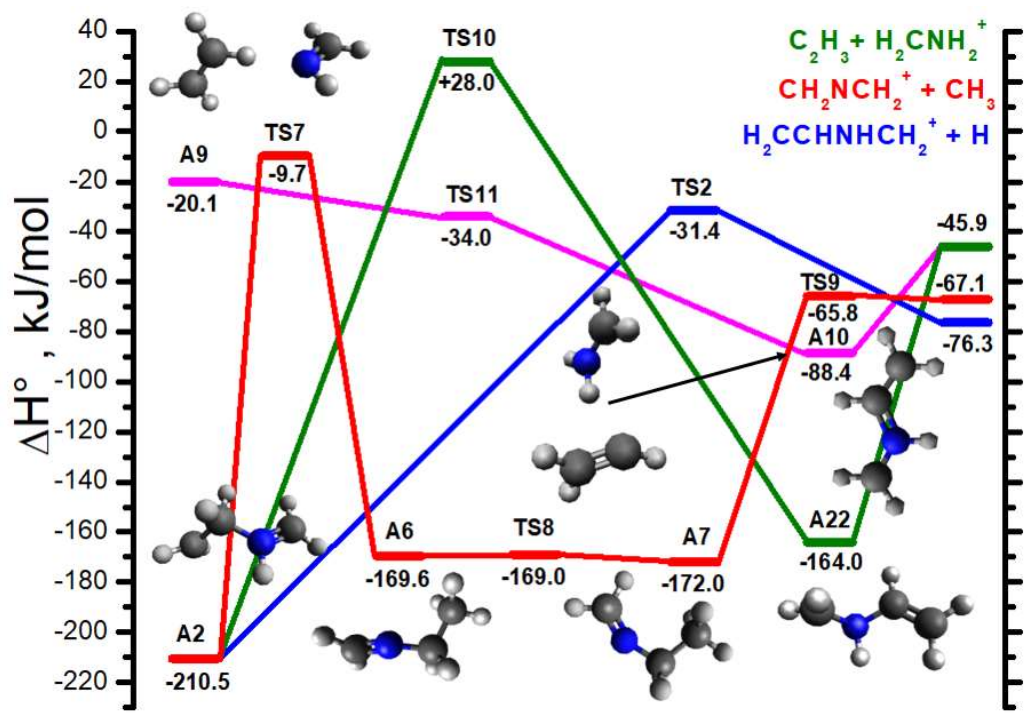

**Fig. 6**. *$H_2CNH^{•+}$ plus $C_2H_4$*: relative enthalpies and reaction pathways leading to products $CH_2CNH_2^+$ plus $CH_3^•$ (in red), $H_2CNH_2^+$ plus $C_2H_3^•$ (in green and pink) and $CH_2CHNHCH_2^+$ plus $H^•$ (in blue). The zero value for enthalpies (in kJ·mol$^{-1}$) corresponds to the separated $H_2CNH^{•+}$ plus $C_2H_4$ reactants. Calculations are at the UCCSD(T)/cc-pVTZ level of theory.

Interestingly, calculations predict different structures for the $C_3H_6N^+$ products from the two isomers. For $HCNH_2^{•+}$ adduct formation proceeds by C-C bond forming, thus leading to the most stable among the $C_3H_6N^+$ isomers, *i.e.* $CH_2CHCHNH_2^+$ (see **Table 1**). On the other hand, the radical character of the N-H terminal in the methanime radical cation leads to C-N bond formation in the intermediate structures, which favours the production of the higher energy $C_3H_6N^+$ isomers c-$CH_2CH_2CHNH^+$ and $CH_2CHNHCH_2^+$. The formation of other isomers with *m/z* 56 cannot be excluded, but a discusson is beyond the scope of this study.

The *m/z* 30 channel can proceed either via the covalently bound adducts or via van der Waals complexes with ethylene, as well as likely exhibiting some contribution from a direct H-stripping at higher collision energies (see the discussion on the H atom transfer reaction with $CH_4$[16]). The collisional energy dependence of the CS is very similar for the two isomers, showing only a slightly more significant decrease with increasing collision energy for $H_2CNH^{•+}$, likely due to the competition with the *m/z* 28 channel, which is not present with the $HCNH_2^{•+}$ isomer.

The *m/z* 28 channel shows the most significant differences between the two isomers. At low collision energies there is a fully-submerged adduct pathway to form this product with both isomers, but for HCNH$_2^{\bullet+}$ this involves significantly more rearrangement and so is kinetically more inhibited, which may explain the much smaller CS for this channel at low collision energies. At high collision energies, the H$_2$CNH$^{\bullet+}$ isomer shows a marked increase followed by a plateau, which is expected to correspond to the slightly endothermic (+37.6 kJ/mol) charge transfer channel, which opens up at high collision energies, thus leading to the observed double-bell effect.

The *m/z* 42 channel is minor but present for both isomers. For the HCNH$_2^{\bullet+}$ ion, there is a barrierless pathway via the adduct, with the small CS being attributed to the fact that it requires multiple rearrangements. For the H$_2$CNH$^{\bullet+}$ ion, the pathway also proceeds barrierlessly via the adduct and is actually the lowest energy pathway though as it also requires a number of rearrangements it is far more kinetically inhibited than the two major channels.

Importantly, present results from the reactivity with C$_2$H$_4$ demonstrate better than reactivity with CH$_4$ (see our accompanying paper [16]) that the two radical cation isomers have markedly different chemistries. This is a strong evidence in support of the selectivity of our ion generation processes and of the fact that reagent ion isomerization prior to reaction is not occurring under our experimental conditions.

## 6. Conclusions

The products of the reaction of the methanimine radical cation (H$_2$CNH$^{\bullet+}$) and its isomer aminomethylene (HCNH$_2^{\bullet+}$) with C$_2$H$_4$ and the dependence of the CSs of their formation pathways on both photon and collision energies have been studied using the CERISES apparatus at the DESIRS beamline. The different behaviour of products clearly show that it is possible to produce the different isomers selectively through the choice of apt precursor molecules. It also shows that isomerisation reactions of the reactants does not occur in the present set-up and conditions.

For both isomers, the channel leading to *m/z* 56 involving a H-elimination of the primarily formed adduct is the main pathway at low collision energies. This implies that the title reactions can serve as intermediate steps to form larger nitrogen-containing species in different ionic environments, including Titan's atmosphere and the structure of products will be influenced by the structure of the cationic isomer. The channel leading to *m/z* 30 (formation of CH$_2$NH$_2^+$) is prominent for both isomers, whereas the product at *m/z* 42 (elimination of a methyl radical from the adduct) is a minor feature, especially in the case of HCNH$_2^{\bullet+}$. All observed channels show a negative dependence of the cross sections on the collision energies at low $E_{CM}$ (smaller than 0.5 eV) which points to exoergic and barrierless processes which are feasible in cold environments like the interstellar medium and Titan's atmosphere.

**Acknowledgements**

We are grateful to the DESIRS beamline team for assistance during the synchrotron measurements and the technical staff of SOLEIL for the smooth running of the facility under projects n◦ 20180118 and 20190249. This work was supported by the European Union's Horizon 2020 research and innovation programme "Astro-Chemistry Origins" (ACO), Grant No 811312. W.G. thanks the Swedish Research Council for a project grant (grant number 2019-04332). M.P. and J.Ž. acknowledge support from the Ministry of Education Youth and Sports of the Czech Republic (grant No. LTC20062). V.R. acknowledges funding for a PhD fellowship from the Dept. Physics, University of Trento.